\documentclass[letter]{aa} 

\usepackage{graphicx}
\usepackage{txfonts}
\usepackage{natbib}
\bibpunct{(}{)}{;}{a}{}{,}
\begin{document} 

   \title{The Gaia-ESO Survey: Galactic evolution of lithium at high metallicity
  \thanks{
 Based on data products from observations made with ESO Telescopes at the La
Silla Paranal Observatory under programmes 188.B-3002, 193.B-0936, and
197.B-1074.}
  }
   \author{S. Randich
          \inst{\ref{oaa}}
      \and L. Pasquini\inst{\ref{eso}}
      \and E. Franciosini\inst{\ref{oaa}}
      \and L. Magrini\inst{\ref{oaa}}
       \and R.J. Jackson\inst{\ref{keele}}
       \and R.D. Jeffries\inst{\ref{keele}}
       \and V. d'Orazi\inst{\ref{padova}}
       \and D. Romano\inst{\ref{bologna}}
        \and N. Sanna\inst{\ref{oaa}}
        \and G. Tautvai{\v s}ien{\. e}\inst{\ref{vilnius}}
       \and M. Tsantaki\inst{\ref{oaa}}
       \and N.J. Wright\inst{\ref{keele}}
          \and G. Gilmore\inst{\ref{ioa}}
          \and T. Bensby\inst{\ref{lund}}
          \and A. Bragaglia\inst{\ref{bologna}}
          \and E. Pancino\inst{\ref{oaa}}
          \and R. Smiljanic\inst{\ref{warsaw}}
          \and A. Bayo\inst{\ref{valp},\ref{valp1}}
          \and G. Carraro\inst{\ref{unipd}}
          \and A. Gonneau\inst{\ref{ioa}}
          \and A. Hourihane\inst{\ref{ioa}}
          \and L. Morbidelli\inst{\ref{oaa}}
          \and C.C Worley\inst{\ref{ioa}}
                }
   \institute{INAF-Osservatorio Astrofisico di Arcetri,
              Largo E. Fermi, 5, 50125 Firenze, Italy\\
              \email{sofia.randich@inaf.it}
              \label{oaa}
         \and European Southern Observatory, Karl Schwarzschild Strasse 2, 85748 Garching, Germany
             \label{eso}
        \and Astrophysics Group, Keele University, Keele, Staffordshire ST5 5BG, United Kingdom 
             \label{keele}
       \and INAF-Osservatorio Astronomico di Padova, Vicolo dell'Osservatorio, 5, 35122 Padova, Italy
       \label{padova}
        \and INAF-Osservatorio di Astrofisica e Scienza dello Spazio, via Gobetti 93/3, 40129, Bologna, Italy
       \label{bologna}
       \and Institute of Theoretical Physics and Astronomy, Vilnius University, Sauletekio av. 3, 10257, Vilnius, Lithuania
       \label{vilnius}
       \and Institute of Astronomy, University of Cambridge, Madingley Road, Cambridge CB3 0HA, United Kingdom 
       \label{ioa}
       \and Lund Observatory, Department of Astronomy and Theoretical Physics, Box 43, SE-221 00 Lund, Sweden
       \label{lund}
       \and Nicolaus Copernicus Astronomical Center, Polish Academy of Sciences, ul. Bartycka 18, 00-716, Warsaw, Poland
       \label{warsaw}
       \and Instituto  de  F\'isica  y Astronom\'ia,  Facultad  de Ciencias,  Universidad de Valpara\'iso, Av. Gran Breta\~na 1111, 5030 Casilla, Valpara\'iso, Chile
       \label{valp}
       \and N\'ucleo Milenio de Formaci\'on Planetaria - NPF, Universidad de Valpara\'iso, Av. Gran Breta\~na 1111, Valpara\'iso, Chile
       \label{valp1}
       \and Dipartimento di Fisica e Astronomia, Universit\`a di Padova, Vicolo dell'Osservatorio 3, 35122 Padova, Italy
       \label{unipd}
             }
   \date{Received ; accepted }

 
  \abstract
   {Reconstructing the Galactic evolution of lithium (Li) is the main tool used to constrain the source(s) of Li enrichment in the Galaxy. Recent results have suggested a decline in Li at supersolar metallicities, which may indicate reduced production.}
   {We exploit the unique characteristics of the Gaia-ESO Survey open star cluster sample to further investigate this issue and to better constrain the evolution of Li at high metallicity.}
   {We trace the evolution of the upper envelope of Li abundance versus metallicity evolution using 18 clusters and considering members that should not have suffered any Li depletion.}
   {At variance with previous claims, we do not find any evidence of a Li decrease at high metallicity. The most metal-rich clusters in the sample ([Fe/H]=$\sim 0.3$) actually show the highest Li abundances, with A(Li)$>$3.4. Our results clearly show that previous findings, which were based on field stars, were affected by selection effects. The metal-rich population in the solar neighbourhood is composed of relatively old and cool stars that have already undergone some Li depletion; hence, their measured Li does not represent the initial interstellar medium abundance, but a lower limit to it.}
   {}


   \maketitle
%

\section{Introduction}

\citet{trimble} summarised  very effectively the relevance of lithium (Li) in modern astrophysics:  `We continue to find it slightly disconcerting that so uncommon an element
as lithium should be so important for studying the structure of outer layers of stars, not to mention
the early Universe. But so it is.' If we were able to fully understand and model Li observations in stars and in the interstellar medium (ISM), we would also be able to answer a number of interesting astrophysical questions.

In particular, focusing on the Galactic evolution of Li abundance and its trend with metallicity, numerous papers have been published since one of the first studies \citep{rebolo}; nevertheless, 
a major question remains open; namely, what Li production sources contribute to the Li enrichment in the Galaxy and can explain the increase from the {plateau value} \citep{spitespite}  observed in metal-poor Population\;{\sc ii} stars to the factor of $\sim 10$ higher Li measured in meteorites and young T Tauri stars \citep[see e.g.][]{dantona, travaglio,cescutti}. \footnote{The {plateau} Li abundance is a factor of about three smaller than the Big Bang abundance 
\citep[e.g. ][]{cyburt, pitrou}; hence, Galactic evolution may actually only need to explain the increase from the Big Bang value to the meteoritic value.}

Whilst it is not easy to summarise and review all the works that have addressed Li abundances and  Galactic Li evolution in the past thirty years, a few recent observational results can be highlighted.  
In particular, we note that amongst the proposed contributors to the Li enrichment in the Galaxy, asymptotic giant branch (AGB) stars ejecta, red giants, supernovae, cosmic ray spallation, and novae, recent observations of novae  have clearly detected  the lines $^7$Li  or $^7$Be (which then decays into $^7$Li) at the early stages, showing that these  systems may represent a dominant source of Li enrichment in the Galaxy \citep{izzo, tajitsu, molaro, izzo18}. Along similar lines, recent observations of large samples of giants provide some support to the idea that these stars may also contribute 
to the Galactic Li enrichment \citep[e.g.][and references therein]{deepack}, as suggested in \citet{romano01}.
Importantly, the empirical evolution of Li with metallicity has been better constrained thanks to the numerous observations and spectroscopic surveys carried out in recent years, and have indeed produced an astonishing amount of optical high-resolution spectra that allow us to study chemical abundances in the different populations of the Galaxy,
investigating simultaneously the evolution of several elements.

Lithium is a fragile element that is destroyed at the relatively low temperature of 2.5$\times 10^6$\; K in stellar interiors, and  may hence be depleted in stellar atmospheres \citep[see e.g.][]{pins}. Therefore, when
looking at the Li versus [Fe/H] distribution, at each metallicity a large dispersion in abundances is observed due to stars that have suffered different amounts of depletion; in order to correctly
define the evolution of the original ISM Li abundance with [Fe/H], it is necessary to  make sure that the upper envelope of the distribution is traced by undepleted stars whose Li content should be representative of the pristine value. 
Based on the observed distribution of Milky Way (MW) field stars in the solar vicinity, several recent studies have suggested that this  upper envelope declines at  supersolar metallicities \citep{delgadomena,ambre,fu,bensbylind,guiglion,stonkute}.
This unexpected result is quite difficult to explain and to model \citep{grisoni}, and
alternative explanations have been proposed. On the one hand, it has been suggested that
the decrease in Li is due to reduced production in the metal-rich regime \citep{prantzos, fu, grisoni}, for example because of  lower AGB yields
and/or  a lower occurrence of nova systems at high metallicity. Alternatively, it has been proposed that
the decline in Li is not real, but rather due to the adopted selection functions of the MW field samples. In particular, \citet{anthony}  point  out how the results of Fu et al. might be affected by selection biases and suggest  that the maximum Li abundance might actually increase at high metallicity \citep[see also][]{cummings1}.
Along similar lines, \citet{guiglion} and \citet{bensby}   speculate  that metal-rich field stars in the solar vicinity are old stars that have migrated from the inner part of the disc, depleting lithium as they travelled and got older. In other words, Li in those stars may not be representative of the original ISM value.

Clearly, Li measurements in young and metal-rich populations, which have presumably not depleted any Li, and the comparison with their more metal-poor counterparts is crucial to discriminating between the two hypotheses.
In this context we exploit the observations of open clusters (OCs) performed by the Gaia-ESO Spectroscopic Survey 
 \citep[GES-][]{GES,GES1}; in particular, the GES OC sample includes several clusters with supersolar metallicity located in the inner Galaxy that are particularly suited to the above purpose.
\section{Sample and lithium determination}
Our study is based on the fifth internal data release of the GES (GESiDR5\footnote{The GESiDR5 catalogue is available for the members of the GES consortium at http:/ges.roe.ac.uk/.}). Spectra for cluster stars were obtained with FLAMES \citep{pasqu02}, either with UVES and the 580 setup or with GIRAFFE and the HR15N setup. Both setups include the Li {\sc i} 6707.8\,\AA\;absorption doublet. 

Lithium abundances in GESiDR5 were derived by different analysis nodes within dedicated working groups
groups (WG10, WG11, WG12; see e.g. Randich et al. 2018 for a detailed description of GES working group data flow and data products). More specifically, one-dimensional (1D), local thermodynamical equilibrium (LTE) abundances (A(Li) -$\log$ N(Li)/N(H)+12) were computed by different nodes adopting GES {recommended} stellar parameters and considering the
Li 6707.8~\AA~feature, either fitting it with spectral synthesis or by measuring  the equivalent width of the line, which was then converted to abundances using a new set of curves of growth specifically derived for GES (Franciosini et al., in prep.). When the Li line was blended with the nearby 6707.4~\AA\,Fe\,{\sc i} line, before computing Li abundances the equivalent widths were corrected for the Fe contribution by using the same grid of synthetic spectra used to derive the curves of growth. The abundances from the different nodes were then combined within each working group and then homogenised to produce the final recommended values (Hourihane et al., in prep.). 

The sample clusters were chosen starting from the list of OCs analysed in GESiDR5. 
We selected mainly young or very young clusters, where at least 4-5 members with supposedly pristine unprocessed Li, representative of the ISM value,  are present. Specifically, the sample includes the following: {\it i.} very young (age $< 100$\;Myr) clusters whose members are pre-main sequence (PMS) or zero age main sequence stars that should have not yet depleted any Li; {\it ii.} clusters older than 100\;Myr, but generally younger than 2\,Gyr, whose upper main sequence (MS) stars are located on the blue (or warm) side of the so-called Li dip \citep[see e.g.][]{boesgaardtripicco,soder93prae,soder93uma,francois,cummings}; no Li depletion is expected for these stars \citep[see e.g.][]{gao} and hence their Li abundance should be representative of the ISM value. 
The only exceptions to these criteria are NGC~2516, NGC~2420, and NGC~2243. The first cluster is slightly older than 100\,Myr, but its members on the blue side of the dip are too bright and were not observed by GES. In the other two clusters, given their ages, stars on the blue side of the dip are no longer on the MS, but are located at the upper turnoff (TO), and they may already have undergone some post-MS Li dilution. Hence the maximum Li that we report for these three clusters is possibly a lower limit to the original ISM value. We decided to retain them as part of the sample in order to enlarge the number of comparison solar-metallicity and metal-poor objects. We also note that some PMS Li depletion may be expected in cool young cluster members due to rotational mixing \citep[see e.g.][]{bouvier}; however,
our average maximum Li abundance for these young clusters is based on stars on the upper envelope of the Li versus effective temperature
(T$_{\rm eff}$) distributions, and we can safely assume that the clusters are sampled well enough that the highest Li stars provide the best approximation to their initial Li. 
The final sample includes 18 OCs, covering the age range between 2~Myr and 5~Gyr; their metallicity (as homogeneously computed by the GES) ranges between [Fe/H]=$-0.38 \pm 0.04$ (NGC~2243) and  [Fe/H]=$+0.26 \pm 0.06$ (Ruprecht~134). 

Cluster membership was obtained following \citet{jackson} and stars with membership probability higher than 80\% were considered. In order to compute the average value of the maximum Li abundance for each cluster, as mentioned, we considered PMS stars in the very young clusters and stars on the blue side of the dip in the older ones. In NGC\,2516 the stars with the highest Li on the red side of the dip were used. Only stars with Li detections (i.e. no upper or lower limits) were taken into account.
The results on the maximum Li are given in Table~\ref{cluster_sample}, along with the associated dispersion. 
\begin{table*}
\caption{Sample clusters, their parameters, and average maximum Li abundance. Ages and metallicities were taken from \citet{spina}, \citet{magrini},
\citet{randich18}, \citet{casali}. The error bars represent the standard deviation of the stars that contribute to the calculation 
of the mean. The actual error in the mean is a factor of $\sqrt{N}$ smaller.}.
\label{cluster_sample}
\begin{tabular}{llrcccc}
\hline\hline
        Cluster  & Age      & R$_\mathrm{GC}$ & [Fe/H] & A(Li)$_\mathrm{max}$ & N stars & T$_{\rm eff}$ range\\
                 & (Gyr)      &  (kpc)     &        &    & used     & (K)  \\
\hline
NGC~6530         &  0.002 & 6.76   & $ -0.041 \pm 0.009$ & $3.38 \pm 0.06$  &  8  & 4490 -- 5300\\ 
Trumpler~14  &  0.002  & 7.62   & $ -0.03 \pm 0.016$ & $3.45 \pm 0.07$ & 62  & 4200 -- 5150\\ 
Chamaeleon~I        &  0.002 & 8.0    & $ -0.07 \pm 0.017$ & $3.25 \pm 0.11$ & 23  & 3475 -- 4310 \\  
$\rho$~Oph   &  0.003 & 8.0    & $ -0.08 \pm 0.006$ & $3.34 \pm 0.14$  & 13  & 3390 -- 4630 \\  
NGC~2264 .   &  0.003  & 8.71   & $ -0.06 \pm 0.04$ & $3.31 \pm 0.07$ & 12  & 4032 -- 5154 \\
IC~4665      &  0.028 & 7.65   & $ 0.00 \pm 0.02$ & $3.32 \pm 0.16$  & 12  & 5240 -- 6250 \\   
IC~2602          &  0.03  & 7.95   & $ -0.02 \pm 0.10$ & $3.29 \pm 0.18$  & 10  & 5290 -- 6705 \\   
NGC~2547         &  0.035 & 8.04   & $ -0.006\pm 0.009$ & $3.27 \pm 0.14$  & 21  & 5615 -- 6930 \\  
NGC~6067     &  0.10  & 6.81   & $  0.20 \pm 0.08$ & $3.41 \pm 0.13$  &  4  & 6630 -- 7370 \\  
NGC~2516         &  0.11  & 7.98   & $ -0.06 \pm 0.05$ & $3.20 \pm 0.07$  &  5  & 6580 -- 6950 \\   
NGC~6259     &  0.21  & 7.03   & $  0.21 \pm 0.04$ & $3.40 \pm 0.06$  &  4  & 6600 -- 7050  \\ 
NGC~6705         &  0.30  & 6.33   & $  0.16 \pm 0.04$ & $3.43 \pm 0.12$  & 10  & 7050 -- 7470 \\   
Berkeley~81      &  1     & 5.49   & $  0.22 \pm 0.07$ & $3.45 \pm 0.15$  & 11  & 6490 -- 7480 \\ 
NGC~6802     &  1.00  & 6.96   & $  0.10 \pm 0.02$ & $3.36 \pm 0.14$  &  5  & 6760 -- 7230 \\   
Ruprecht~134 &  1.00  & 4.60   & $  0.26 \pm 0.06$ & $3.42 \pm 0.10$ & 11  & 6500 -- 6990 \\  
Trumpler~20  &  1.4   & 6.86   & $  0.15 \pm 0.07$ & $3.38 \pm 0.13$  &  4  & 7025 -- 7170\\  
NGC~2420     &  2.20  & 10.76  & $ -0.13 \pm 0.04$ & $3.05 \pm 0.07$  & 21  & 6290 -- 6550 (upper TO)\\  
NGC~2243         &  4.9   & 10.4   & $ -0.38 \pm 0.04$ & $2.96 \pm 0.06$  &  5  & 5890 -- 6190 (upper TO)\\
             &        &        &                   &                  &     & \\ \hline
\end{tabular}
\end{table*}
\section{Results and discussion}
\begin{figure*}
\centering
   \resizebox{\hsize}{!}{\includegraphics{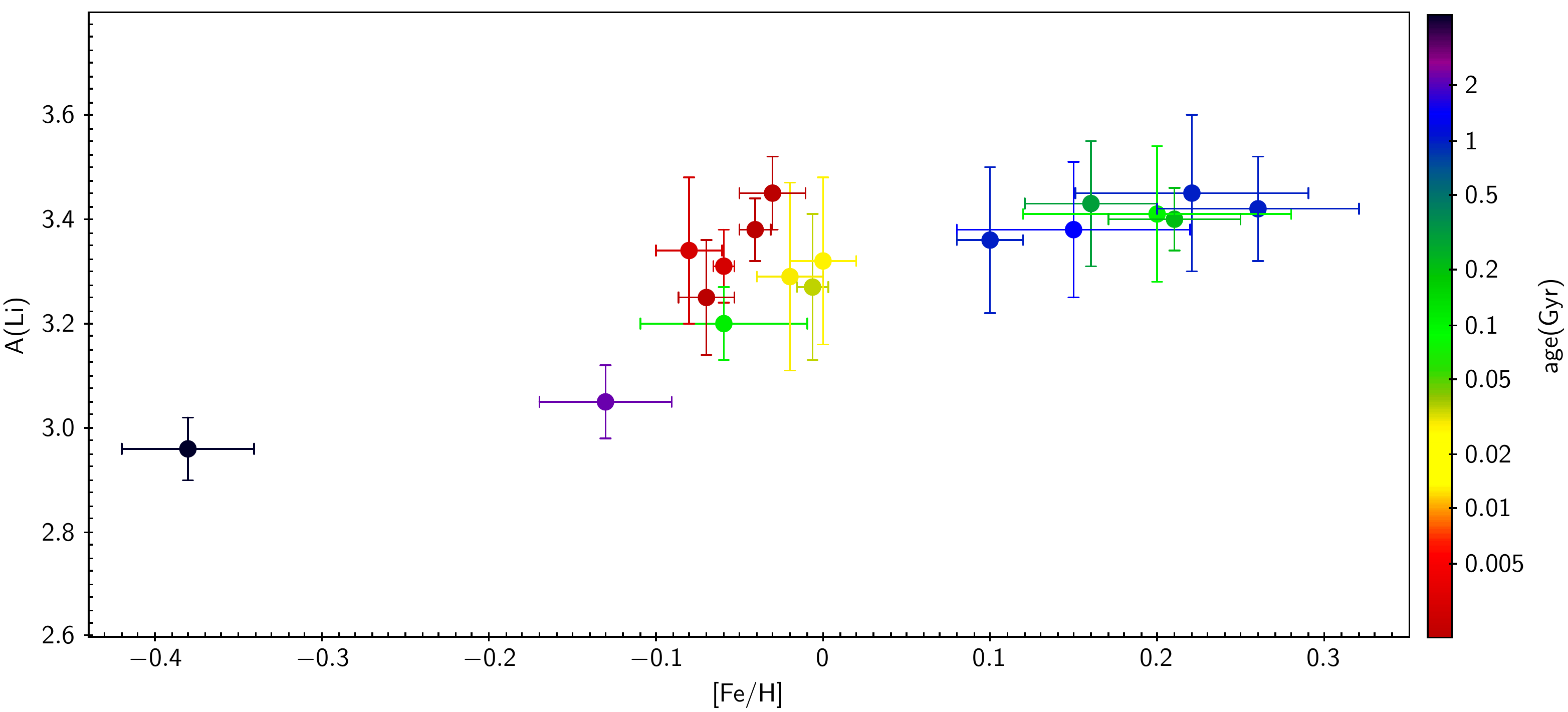}}
           \caption{Average maximum Li abundance (see text) as a function of the cluster metallicity. Clusters are colour-coded by age. }
              \label{Fig_ev}%
    \end{figure*}

Figure~\ref{Fig_ev} summarises our results for the evolution of Li as a function of metallicity. The figure clearly suggests, at variance with previous claims, that the maximum Li abundance does not decrease at high metallicities, at least up to about [Fe/H]\;$\sim +0.3$. Actually, the two clusters with subsolar metallicities show a maximum Li abundance below A(Li)=3.1 (but, as mentioned, stars in these clusters may have already suffered some Li dilution),  while all the others are above or close to 3.3, and the two most metal-rich (Ruprecht~134 and Berkeley~81) have A(Li)$_\mathrm{max}\sim 3.4$, which is also much higher than the Li abundances measured in field stars at similar metallicities.
Therefore, the main conclusion of this study is that we do not see any evidence of a decrease in Li at supersolar metallicity. Rather, the opposite might be true, and the data may suggest a positive Li versus [Fe/H] trend, in agreement with the suggestions of \cite{anthony}. 
Whilst a Bayesian analysis also supports a positive Li versus Fe correlation, 
this needs to be confirmed with a larger number of clusters, in particular young ones in the metal-poor regime. 

\begin{figure}
 \centering
   \resizebox{\hsize}{!}{\includegraphics{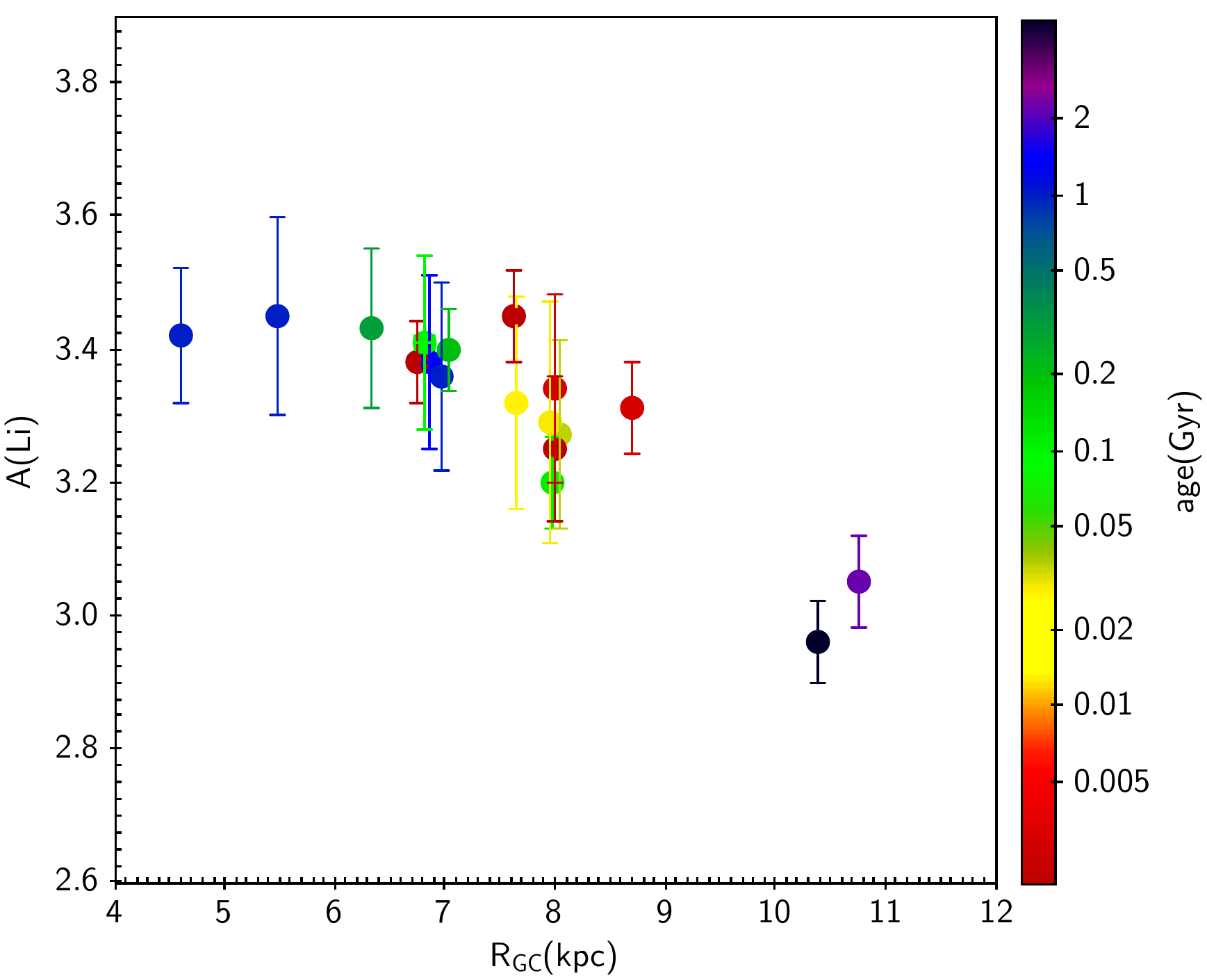}}
           \caption{Same as Fig.\,\ref{Fig_ev}, but lithium is plotted as a function of the cluster Galactocentric distance.}
              \label{Fig_grad}%
    \end{figure}

Similarly, the data would suggest a mild trend with Galactocentric distance (R$_{GC}$, see  Fig.~\ref{Fig_grad}) or a shallow gradient. To our knowledge this is the first time that a Li gradient has been shown observationally; however, also in this case, a larger cluster sample is needed to derive firm conclusions and to perform a comparison with the models.

Given that several recent studies have claimed  Li decrease at high metallicity, the opposite of our results, we investigate possible reasons for the disagreement in the following.
\subsection{Three-dimensional and non-LTE corrections}
Lithium abundances from GES have been computed using LTE and plane-parallel atmospheres, while a more detailed analysis would require considering three-dimensional (3D) models and non-LTE effects \citep[NLTE, e.g.][]{lind,klevas,harut}.
   \citet{ambre} and \citet{fu} both applied NLTE corrections using \citet{lind} computations, still with 1D models. Unfortunately, we cannot derive star-by-star NLTE abundances for the whole sample since a detailed grid of NLTE + 3D corrections covering the parameter space of our sample stars is not available in the literature. However, we  used the tool provided by \citet{harut}\footnote{https://pages.aip.de/li67nlte3d} to estimate the effect of NLTE + 3D corrections. Specifically, we fixed the value of the Li abundance (A(Li)$=2.7$ -- the maximum available in their grid), and changed the star temperature and metallicity. As already described in the original paper, the corrections decrease with increasing effective temperature and, for a fixed temperature (e.g. 6500~K), they slightly increase for higher metallicity. More importantly, the corrections are positive (i.e. by applying 3D+NLTE corrections   higher Li abundances are obtained).
For example, at solar metallicity we have A(Li)$_\mathrm{3D-NLTE}-$A(Li)$_\mathrm{1D-LTE}=\Delta$A(Li)\,= 0.055 and 0.024\,dex for T$_\mathrm{eff}\,$=\,6000 and 6500~K, respectively; at [Fe/H]$=0.3$ we have $\Delta$A(Li)\,=\,0.076 and 0.038 for the same effective temperatures. For a given effective temperature the dependence of the corrections on A(Li) is negligible. Since the members of the metal-rich clusters used to infer the maximum Li abundance are all warmer than 6500~K, we conclude that the NLTE-corrections should be well below 0.1~dex, and of positive sign. Neglecting NLTE effects is thus not at the root of the discrepancy between our results for the open clusters and the field stars from other studies.
\subsection{Sampling different populations}
\begin{figure}
   \centering
   \resizebox{\hsize}{!}{\includegraphics{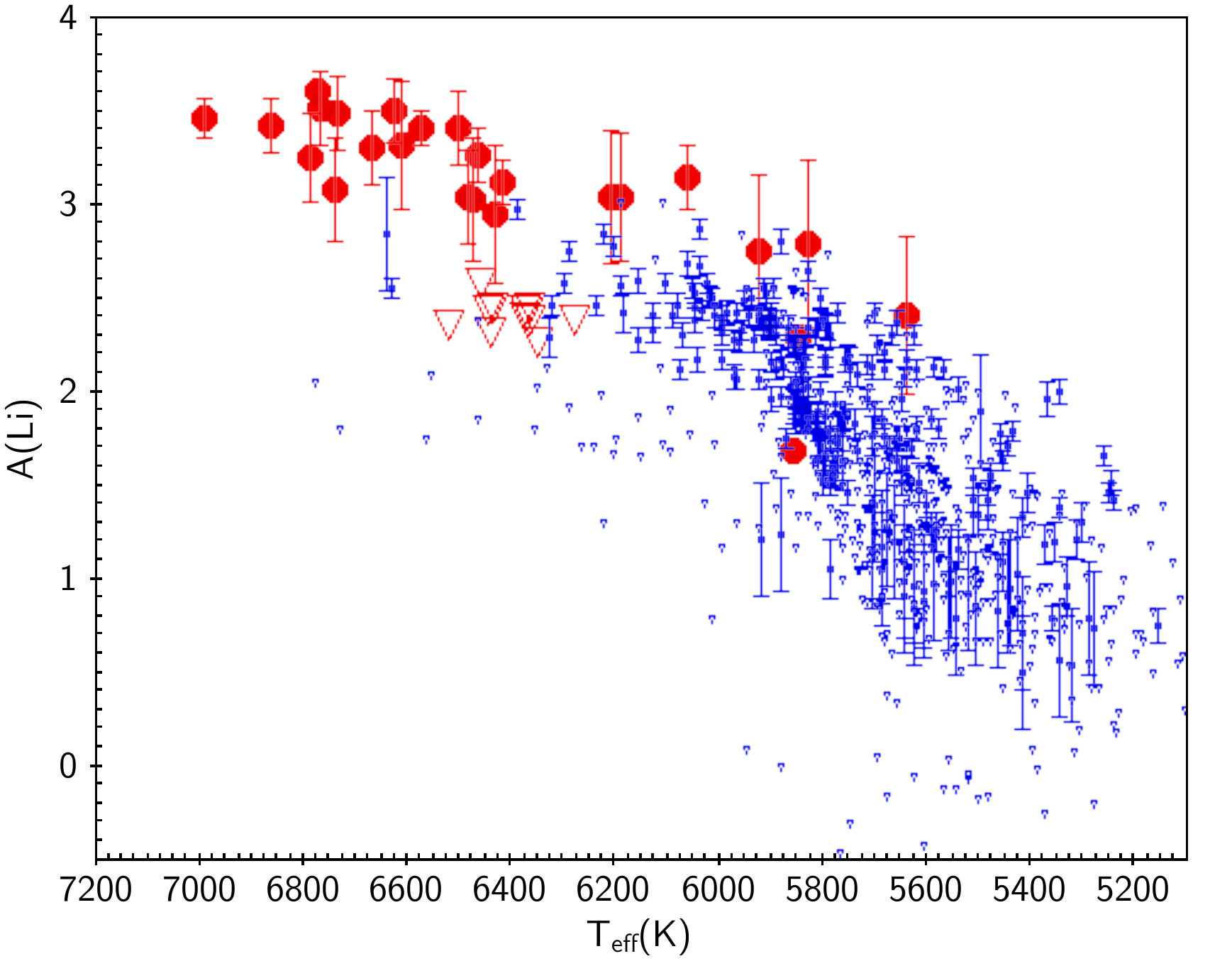}}
           \caption{Lithium abundance as a function of effective temperature for metal-rich ([Fe/H]$>0$) GES
           MW field stars (blue points) and Ruprecht~134 (red points). Circles and triangles indicate Li detections and upper limits, respectively. We do not indicate T$_{\rm eff}$ errors in order to have a clean plot. For  the MW and for Ruprecht\;134 the errors are of the order of 50-100\;K.}
              \label{Fig_rup-mw}%
    \end{figure}

Most spectroscopic surveys of MW field stars are affected by selection criteria, such as the star's distance (limiting magnitude) or colour selection, that may introduce hidden biases.
Observing OCs allows us to eliminate these biases because they are selected based on their combinations of age and metallicity, irrespective of their distance. In addition, since the
stars are members of the clusters, we are able to know exactly their evolutionary status and physical parameters, and to cover a wide range of  evolutionary stages and masses. In our  case we have the additional advantage that the Li dip in many clusters is well defined, so we can use it to guide our analysis. 
When looking at the main differences between cluster and the field samples observed in spectroscopic surveys, two aspects are important.
First, the samples of field stars  \citep[see e.g.][]{ambre,fu} 
contain in the high-metallicity bin only MS stars with effective temperatures below $\sim$6400 K (i.e.  stars located on the cool side of the Li dip). The highest Li values in our sample are instead found amongst stars on the hot side of the dip. This is clearly seen in Fig.~\ref{Fig_rup-mw}, where we compare Ruprecht\;134 with the GES MW stars with available Li measurements.
The Ruprecht\;134 data points show the typical behaviour of Li versus temperature, with the Li dip at around 6400 K, the stars on the hot side with the highest Li values (likely not depleted), and members on the cool side of the dip already showing  temperature-dependent depletion even for such a relatively young age. 
Second,  whilst for the GES and AMBRE surveys neither Fu et al. nor Guiglion et al.  provide  an age distribution for the stars in their samples, we  note that for GES  \citet{thompson} demonstrated that the adopted selection criteria penalise stars younger than 2~Gyr, and that the  sample is biased towards stars with ages between 2 and 5~Gyr. We  therefore have evidence that these surveys observe  metal-rich stars in the field that are cooler than the dip and older than 1~Gyr, and have hence undergone Li depletion. Our maximum Li value for the metal-rich stars is indeed substantially higher (a factor of two-three) than the values observed among field stars surveys.  We therefore conclude that with OCs we sample a young hot population of metal-rich stars that is not sampled by present surveys of the MW field. This population does not show any decrease in Li with respect to solar-metallicity stars.
\section{Conclusions}
In this work we assume that the Li ISM at a given metallicity is well represented by the average Li abundance of OC members warmer than the Li dip, or PMS stars in very young clusters. Based on a sample of 18 OCs, we find that clusters with [Fe/H] $< -0.1$ show a  maximum Li$\sim$3.1,
while the solar- and supersolar-metallicity stars all have   higher values, which peak at A(Li)$\sim 3.4$ for the two most metal-rich objects. A shallow Li gradient may also be present.
Therefore, our results do not support the claim that Li decreases in the ISM for high metallicity, at least up to [Fe/H]\,=\,0.3. On the contrary, if anything a mild increase may be present.

We suggest that the discrepancy between our results and those of other studies can be fully explained because spectroscopic surveys observed old metal-rich MW field stars on the cool side of the Li dip that already suffer noticeable MS Li depletion  when a few hundred million years old. Since the observed metal-rich stars tend to be relatively old, their highest Li abundance is 0.2$-$0.3~dex lower than the original value, and can be considered only as a loose lower limit to the ISM Li abundance. 
In other words, our results strongly indicate that the observed decrease in Li for metal-rich field stars is not `real'; rather, it is due to  stellar evolution and lithium depletion mechanisms, and it is enhanced by sample selection effects.
\begin{acknowledgements}
These data products
have been processed by the Cambridge Astronomy Survey Unit (CASU) at the
Institute of Astronomy, University of Cambridge, and by the FLAMES/UVES
reduction team at INAF/Osservatorio Astrofisico di Arcetri. These data have
been obtained from the Gaia-ESO Survey Data Archive, prepared and hosted by
the Wide Field Astronomy Unit, Institute for Astronomy, University of
Edinburgh, which is funded by the UK Science and Technology Facilities
Council. This work was partly supported by the European Union FP7 programme
through ERC grant number 320360 and by the Leverhulme Trust through grant
RPG-2012-541. We acknowledge the support from INAF and Ministero
dell'Istruzione, dell'Universit\`a e della Ricerca (MIUR) in the form of the
grant "Premiale VLT 2012", PRIN-INAF 2014, Premiale MITiC. The results presented here benefit from
discussions held during the Gaia-ESO workshops and conferences supported by
the ESF (European Science Foundation) through the GREAT Research Network
Programme.
TB was partly funded by the grand 2018-04857 from the Swedish Research Council, and partly by the project grant ’The New Milky Way’ from the Knut and Alice Wallenberg Foundation.
\end{acknowledgements}

%
%
\bibliographystyle{aa}
\bibliography{srandich}
\end{document}